\documentclass[aps, prl, showpacs, preprintnumbers, twocolumn, floatfix]{revtex4-1}
\usepackage{graphicx}
\usepackage{hyperref}
\usepackage{mathptmx, textcomp}
\usepackage[latin1]{inputenc}

\usepackage{xcolor}
\hypersetup{
    colorlinks,
    linkcolor={red!50!black},
    citecolor={blue!50!black},
    urlcolor={blue!80!black}
}

\usepackage{changes}

\newcommand{\Barium}{$^{138}\textrm{Ba}^+$ }
\newcommand{\Ba}{Ba$^+$ }
\newcommand{\Rubidium}{$^{87}\textrm{Rb}$ }
\newcommand{\Rbplus}{Rb$^+$ }

\newcommand{\commentOut}[1]{}

\begin{document}
\title{Reactive two-body and three-body collisions of \Ba in an ultracold Rb gas}
\author{Artjom Kr\"ukow}
\author{Amir Mohammadi}
\author{Arne H\"arter}
\author{Johannes Hecker Denschlag}
\affiliation{Institut f\"{u}r Quantenmaterie and Center for Integrated Quantum Science and
Technology IQ$^{ST}$, Universit\"{a}t Ulm, 89069 Ulm, Germany}

\date{\today}

\begin{abstract}
We analyze reactive  collisions of a single \Ba ion in contact with an ultracold gas of  Rb atoms at low three-body collision energies of 2.2(9) mK$\times k_{\mathrm{B}}$.
Mapping out the \Ba loss rate dependence on the Rb atom density we can discern two-body reactive collisions from three-body ones and for the first time determine both rate coefficients which are  $k_2=3.1(6)(6)\times 10^{-13}\textrm{cm}^{3}\textrm{s}^{-1}$ and $k_3=1.04(4)(45)\times 10^{-24}\textrm{cm}^{6}\textrm{s}^{-1}$,  respectively (statistical and systematic errors in parenthesis).
Thus, the measured ternary recombination dominates over binary reactions  even at moderate atom densities of $n\approx 10^{12}\: \textrm{cm}^{-3}$.
The results for \Ba and Rb are representative for a wide range of cold ion-atom systems and can serve as a guidance for the future development of the field of hybrid atom-ion research.
\end{abstract}

\pacs{34.50.-s, 34.50.Lf, 37.10.-x,}

\maketitle

Cold atom-ion physics in hybrid traps is a young, developing field \cite{Haerter2014,Sias2014,Willitsch2014}, which builds on the relatively long-range $r^{-4}$ polarization potential between atom and ion.
In general, this potential promises large cross sections and therefore strong interactions between particles.
As a consequence, a number of interesting research proposals have been brought forward  ranging from sympathetic cooling down to ultracold temperatures \cite{Krych2011},   to studying the physics of strongly correlated many-body systems, \textit{e.g.,} ultracold charge transport \cite{Cote2000}, novel many-body bound states \cite{Cote-2002} and strong-coupling polarons \cite{Casteels-2011},  quantum information processing \cite{Doerk-2010}  and quantum simulation  \cite{Bissbort-2013}. 
Most of these ideas rely on interactions mediated by elastic atom-ion collisions, while inelastic collisions and chemical reactions are undesired as they represent a time limit for the suggested experiments.
Therefore it is important to identify and investigate possible reactions and to eventually gain control over them.
Inelastic processes can be divided up into classes such as two-body or three-body collisions.
In general, binary collisions are dominant at low enough atomic densities, while ternary collisions will eventually take over with increasing density.
This knowledge has been extensively applied in the field of ultracold neutral atoms by typically working with low enough atomic densities (e.g. smaller than about 10$^{14}$cm$^{-3}$ for $^{87}$Rb)  in order to keep three-body losses negligible \cite{ketterle1999}.
Considering the low-density limit, theoretical predictions for cold hybrid atom-ion systems  have been focussing on binary inelastic/reactive atom-ion collisions (e.g.~\cite{Cote2000PRA,DaSilva2015}) which were discussed  as the limiting factors for proposed atom-ion experiments \cite{Makarov-2003,Krych2011, Sullivan-2012, Tomza-2015}.
Along the same lines, measurements on atom-ion reactions in the low mK regime were, until recently,  unanimously interpreted in terms of pure two-body decay \cite{Grier-2009, Zipkes2010PRL,Schmid2010, Hall2011, Smith2013, Haze2015}.

In this Letter we show, however, that in general the decay analysis requires simultaneous consideration of both two- and three-body reactions.
Our measurements reveal that at mK temperatures inelastic three-body collisions of the ion can dominate over its two-body reactions, even at moderate atomic densities down to 3$\times$10$^{11}$cm$^{-3}$.
Indeed, the main focus of this work lies on how to clearly distinguish  two-body from  three-body processes and extract the corresponding rate  coefficients.
One could in principle argue that in order to study only two-body reactions the atomic density simply needs to be lowered sufficiently.
This is, however, not practical in standard set-ups with magnetic or dipole traps because the resulting reaction rate can be so small that the ion lifetime exceeds the atomic cloud lifetime.
Alternatively, one could consider working with a magneto-optical trap (MOT) which allows for both, low densities and long lifetimes due to continuous loading.
However, in a MOT the reaction rate measurements of the ground state atoms are swamped beneath a background of reactions of electronically excited atoms occurring at much higher rates.

For our investigations we use a heteronuclear combination of Ba$^+$-Rb, where both two-body and three-body collisions lead to reactions and hence  to ion loss in the experiment. 
This complements a recent experiment of ours  with homonuclear Rb$^+$-Rb-Rb  \cite{Haerter-2012} for which reactive and inelastic two-body collisions are either forbidden or irrelevant.
Furthermore, we note that in parallel to the work discussed here, we have studied the energy scaling of atom-ion three-body recombination \cite{Kruekow2016}.

We measure the density dependence of the reaction rate $\Gamma=k_2\cdot n + k_3\cdot n^2$ and extract the binary and ternary loss rate coefficients $k_2$ and $k_3$.
Here, $n$ is the peak atom density at the cloud center where the ion is located.
For the analysis  the evolution of $n(t)$ needs to be included, as the atom cloud is decaying during the time $t$ due to elastic atom-ion collisions.
We experimentally determine  $n(t)$ by excluding experimental runs where the ion has undergone a reaction during the interaction time $t$ in order to avoid systematic errors introduced by reactive collisions.

The experiments are performed in a hybrid apparatus that has already been described in detail elsewhere \cite{Schmid2012}.
We prepare a single \Barium in a linear Paul trap and bring it into contact with an ultracold cloud of spin polarized \Rubidium ($F=1, m_F=-1$).
The atoms are prepared at a separate location from which they are  transported to the Paul trap and loaded into a far off-resonant crossed optical dipole trap.
During the final preparation stage for the atoms, the cloud and the ion are separated by about $100\:\mu$m along the Paul trap axis to avoid unwanted atom-ion interaction.
By ramping one endcap voltage of the linear Paul trap to its final value,  we shift the ion into the center of the atom cloud within $10\:$ms and start the atom-ion interaction.
We use thermal atom clouds consisting of typically $N \approx 40 - 135 \times 10^3$ atoms at temperatures of $T \approx 330\:$nK with peak densities between $n \approx 6 \textrm{ to } 84 \times 10^{11}\:\textrm{cm}^{-3}$.
The $\textrm{Ba}^+$ ion is confined in a linear Paul trap which is operated at a frequency of $4.21\:\textrm{MHz}$ with radial and axial trapping frequencies of $(\omega_r; \omega_a) =2\pi \times (59.5;38.4)\:\textrm{kHz}$.
Single $^{138}$Ba$^+$ ions are loaded by isotope selective, resonant two-photon ionization.
Using standard laser cooling techniques these are cooled to Doppler temperatures of $\approx 0.5\:$mK.
Before immersing the \Ba into the atomic bath we switch off the laser cooling, which guarantees  that the ion is in the electronic ground state during the atom-ion interaction.
The average kinetic energy $\overline{E}_{\mathrm{kin}}$ of the ion is determined by the interplay of elastic collisions and the driven micromotion \cite{Berkeland1998, Devoe2009, Zipkes2010PRL, Cetina2012, Krych2013}. $\overline{E}_{\mathrm{kin}}$ is adjusted by tuning the excess micromotion of the ion and sets the average three-body collision energy $\overline{E}_{\mathrm{col}}$, through the relation $\overline{E}_{\mathrm{col}}\approx 0.55 \ \overline{E}_{\mathrm{kin}}$ \cite{Kruekow2016}.
For the experiments discussed in the following we work either at an energy of $\overline{E}_{\mathrm{kin}}\approx$ 4 or of 70 mK$\times k_{\mathrm{B}}$.

\begin{figure}
\includegraphics[width=\columnwidth]{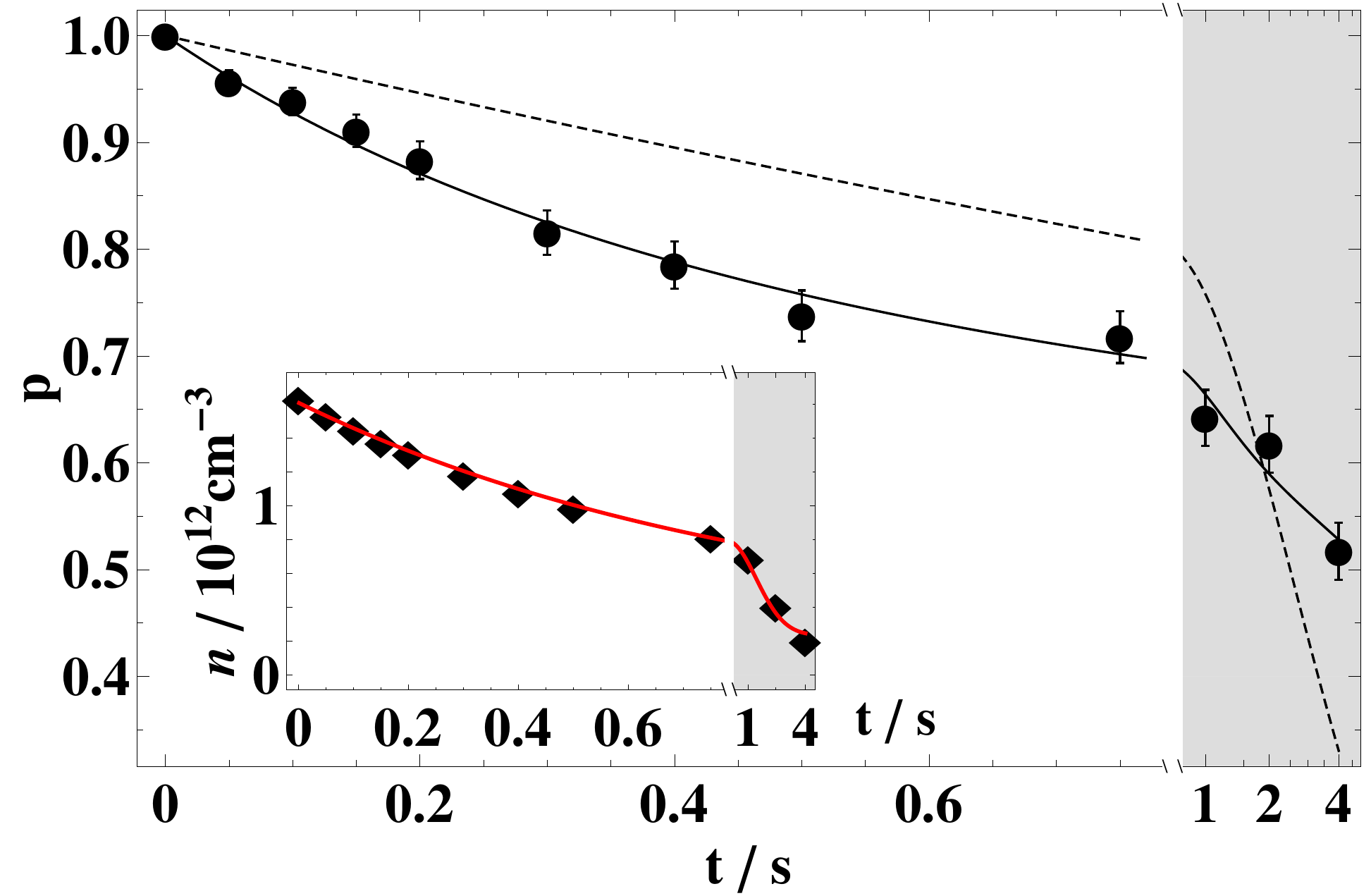}
\caption{(color online) Probability $p$ to detect the \Ba ion after the interaction time $t$ with a Rb atom cloud at an average ion kinetic energy of $\overline{E}_{\mathrm{kin}} \approx 67\:$mK$\times k_{\mathrm{B}}$.
One data point is the average of the binary result over roughly 330 single ion experiments.
A fit (solid line) taking into account the decay of the atom density during the interaction, reproduces this behavior while a simple exponential fit (dashed line) does not. The inset shows the corresponding atom density evolution, which is well described by an exponential decay with an offset (solid line).
Note the time scale change at 0.8$\:$ s as indicated by the shaded background.
All error bars represent the $1\sigma$ statistical uncertainty of the measurements.
}
\label{fig1}
\end{figure}

We start our investigations by measuring the lifetime of a single \Ba in contact with a thermal cloud of Rb atoms.
For this, we immerse the ion ($\overline{E}_{kin}\approx 67\:$mK$\times k_{\mathrm{B}}$) into the atom cloud  (density $n \approx 16 \times 10^{11}\:\textrm{cm}^{-3}$) for various periods of time $t$.
 After the interaction, we check if the \Ba is still present by switching on the laser cooling for $100\:$ms and collecting its  fluorescence on a EMCCD camera.
If no \Ba is detected, we conclude that a reaction must have taken place.
If we apply additional laser cooling with a red-detuning of 2 GHz for several seconds,  typically 50\% of the initially not detected \Ba ions can be recovered.
These ions have gained  high kinetic energies in a chemical reaction, which will be discussed later.
Figure \ref{fig1} shows the measured probability $p$ to detect the \Ba ion as a function of the interaction time $t$ (please note the time scale change after $0.8\:$s).
We model the decay  using the rate equation $\dot{p}=- \Gamma (t) \cdot p$, with the loss rate $\Gamma(t)=k_2\cdot n(t) + k_3\cdot n(t)^2$, where $n(t)$ is the time dependent atom density  at the location of the ion.
Integrating the equation yields
\begin{equation}
\label{eqn1}
p(t)=\exp (-\int^{t}_{0}\Gamma(t') dt').
\end{equation}
A constant density $n(t)$ would lead to an exponential decay,  $p(t)=\exp (- \Gamma t)$, which does not describe the observed loss very well (Fig. \ref{fig1} dashed line).
As the inset of Fig. \ref{fig1} shows the density decreases during the interaction time.
This is because elastic atom-ion collisions either remove atoms from the shallow atom trap or heat up the atomic ensemble.
If we take into account the decay of $n(t)$ a fit of equation \ref{eqn1} (solid line)  describes the data very well.
\begin{figure}
\includegraphics[width=\columnwidth]{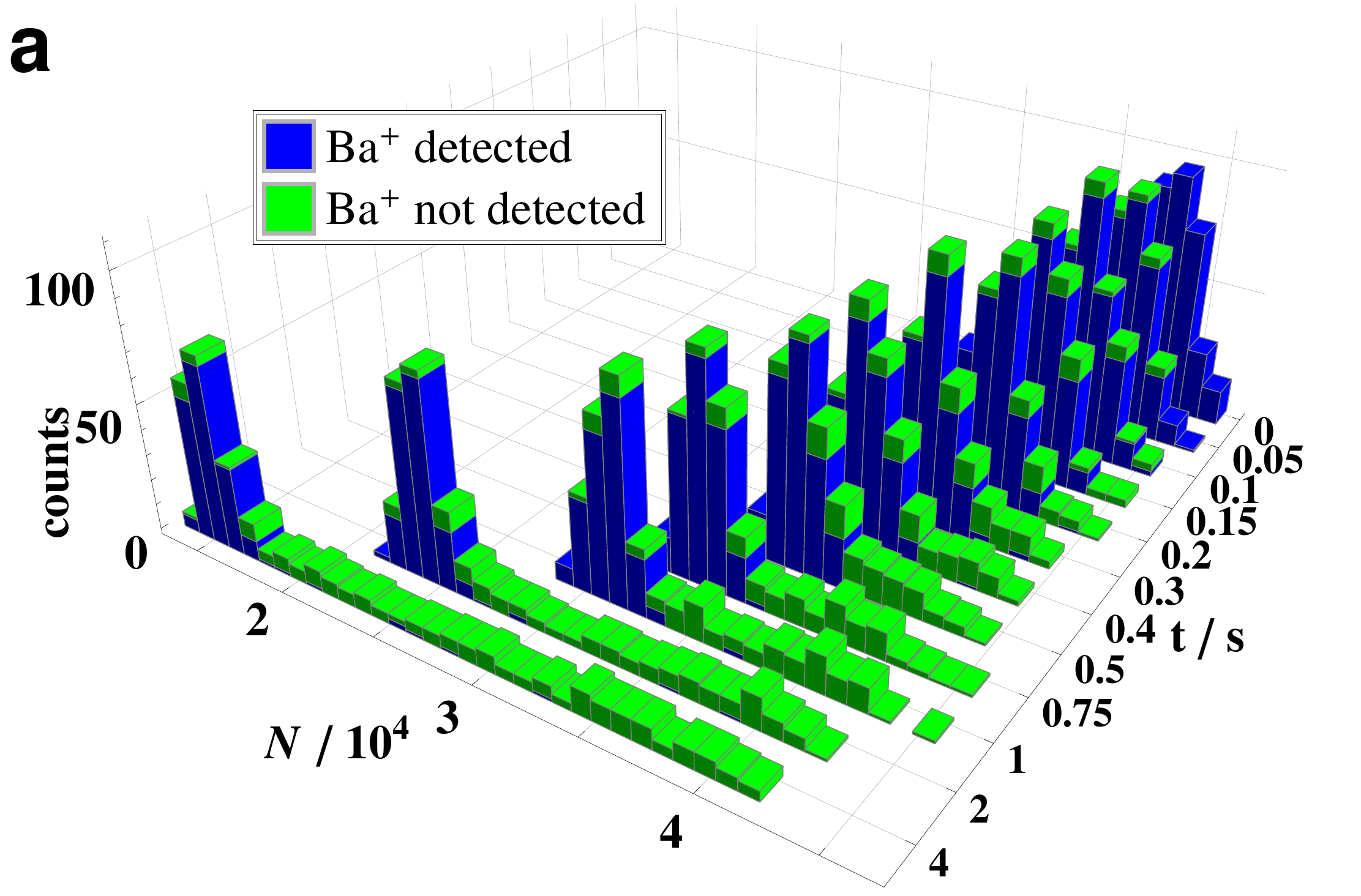}
\includegraphics[width=\columnwidth]{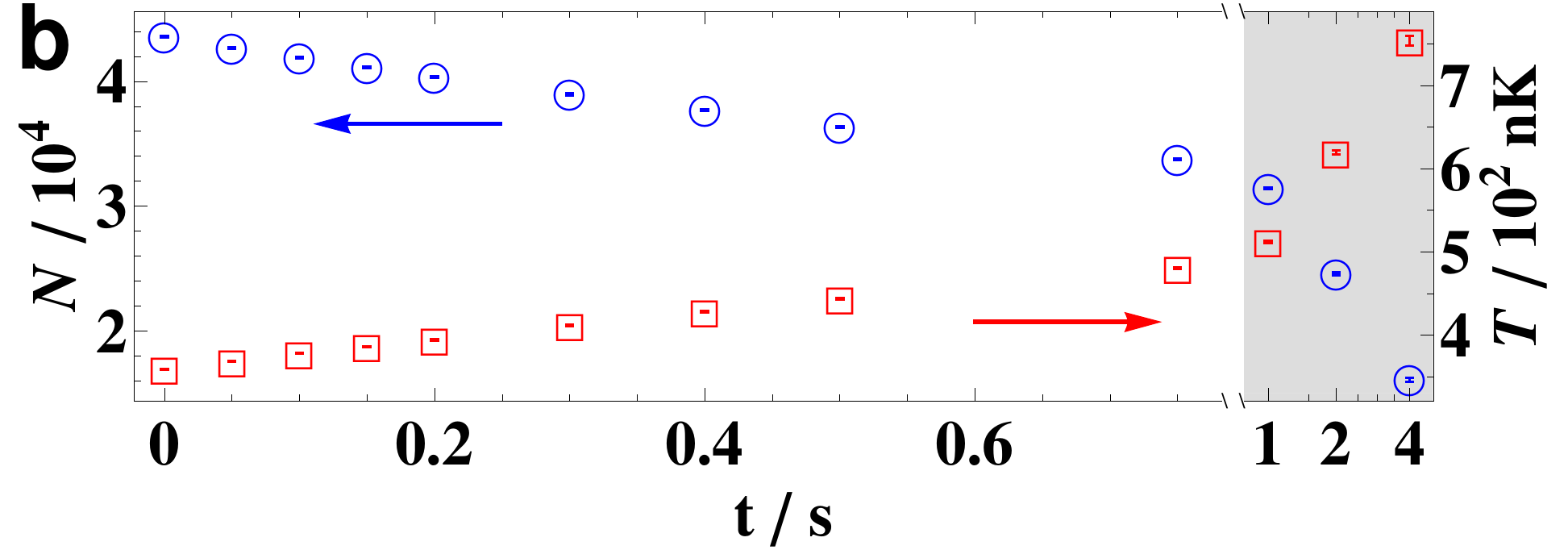}
\caption{(color online) \textbf{(a)} Histogram of the atom numbers $N$ belonging to the measurement in Fig. \ref{fig1}. A Gaussian atom number distribution develops a broad tail with increasing interaction times. Experimental runs where the \Ba ion was detected (not detected) after the interaction are marked in blue (green), respectively. Atom numbers within the tail (Gaussian peak) of the distribution correspond to runs with (without) a reactive collision, respectively. \textbf{(b)} Average atom number $N$ (circles) and temperature $T$ (squares) over all runs without reactions, corresponding to the blue colored counts in panel (a).
}
\label{fig2}
\end{figure}
In other words, for a proper description of the ion loss $\Gamma$ and to determine the rate constants $k_2$ and $k_3$, the evolution of the density $n(t)$  has to be accurately determined.
This, however, is somewhat involved and will be discussed in the following.

To determine $n(t)$, we measure the remaining atom number $N$ and temperature $T$ of the cloud via absorption imaging after $15\:$ms time of flight.
Figure \ref{fig2}a shows histograms of the atom number distributions for various interaction times $t$.
Initially the distribution is Gaussian. As time goes on, elastic atom-ion collisions shift this  distribution towards lower atom numbers.
In addition, a broad tail develops. This tail can be explained as a consequence of reactive atom-ion collisions that release substantial amounts of energy which ejects the product ion out of the atom cloud onto a large orbit trajectory in the Paul trap.
This is consistent with the recovery of hot \Ba ions when additional far red-detuned laser cooling is applied, as mentioned in the previous paragraph.
Although two-body and three-body reactions at mK temperatures are predicted to dominantly produce translationally cold molecular BaRb$^+$ ions \cite{ Hall2013MP,JPR-2015}, additional kinetic energy can be released in fast secondary processes such as photo-dissociation or collisional relaxation.
Once the ion is on the large orbit, the atom-ion collision rate is significantly reduced, essentially stopping the continuous atom loss \cite{Haerter-2012}.
From Fig. \ref{fig2}a we find that the counts in the tail almost exclusively correspond to experimental runs where a reaction with \Ba occurred (green color), whereas the counts in the Gaussian dominantly correspond to runs without reactions (blue color).

For our analysis we only consider system trajectories without reactions, making sure that the ion has been constantly exposed to the central density $n(t)$.
The average atom number $N$ and temperature $T$ of these post-selected trajectories are plotted in Fig. \ref{fig2}b.
We then  calculate the peak atom density (shown in Fig. \ref{fig1} inset) $n=(\frac{m}{2\pi k_{\mathrm{B}}})^{3/2} \cdot \frac{\omega_x \omega_y \omega_z N}{T^{3/2}}$, with the mass $m$, using separately measured trap frequencies $(\omega_x , \omega_y , \omega_z)$ of the atom dipole trap.
From these sampling points we extract the time dependent density $n(t)$.

\begin{figure}
\includegraphics[width=\columnwidth]{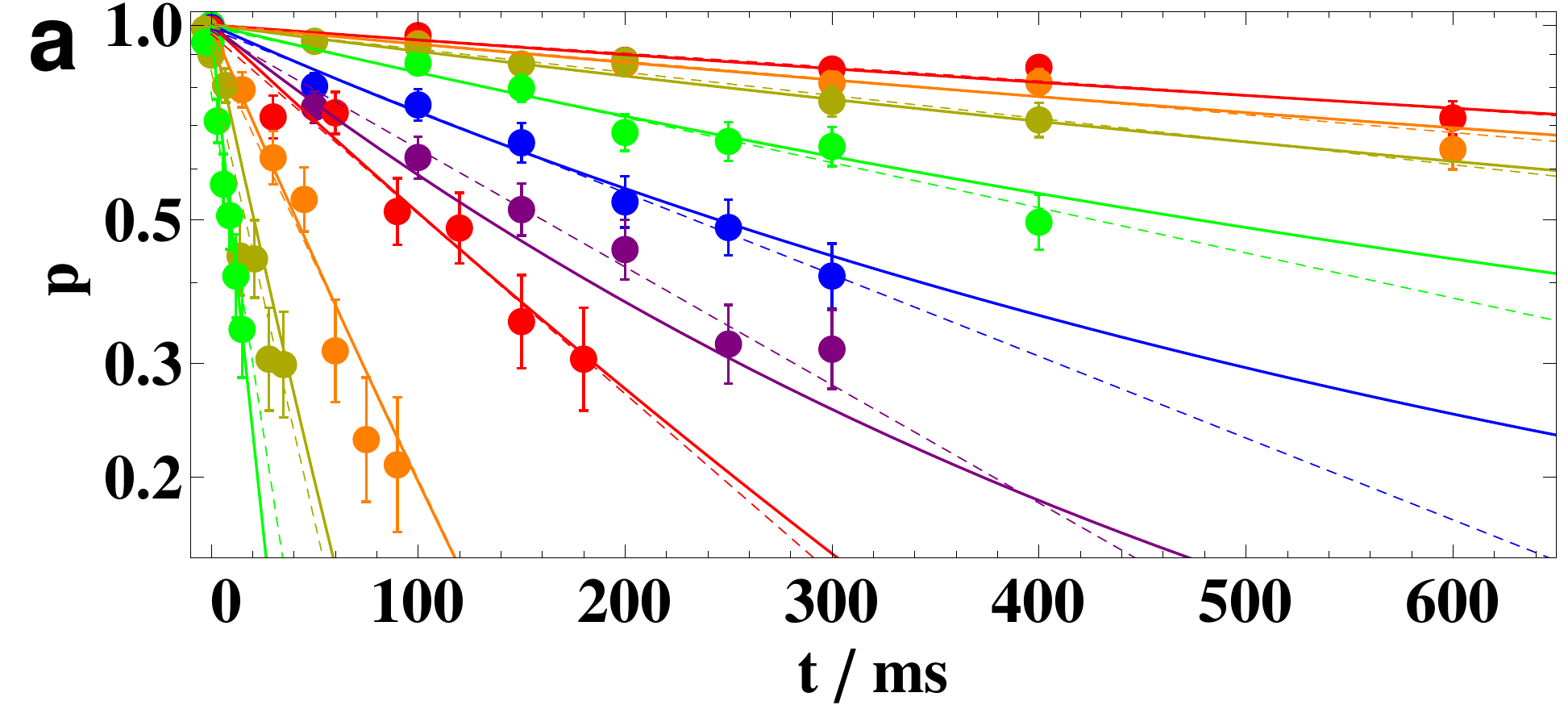}
\includegraphics[width=\columnwidth]{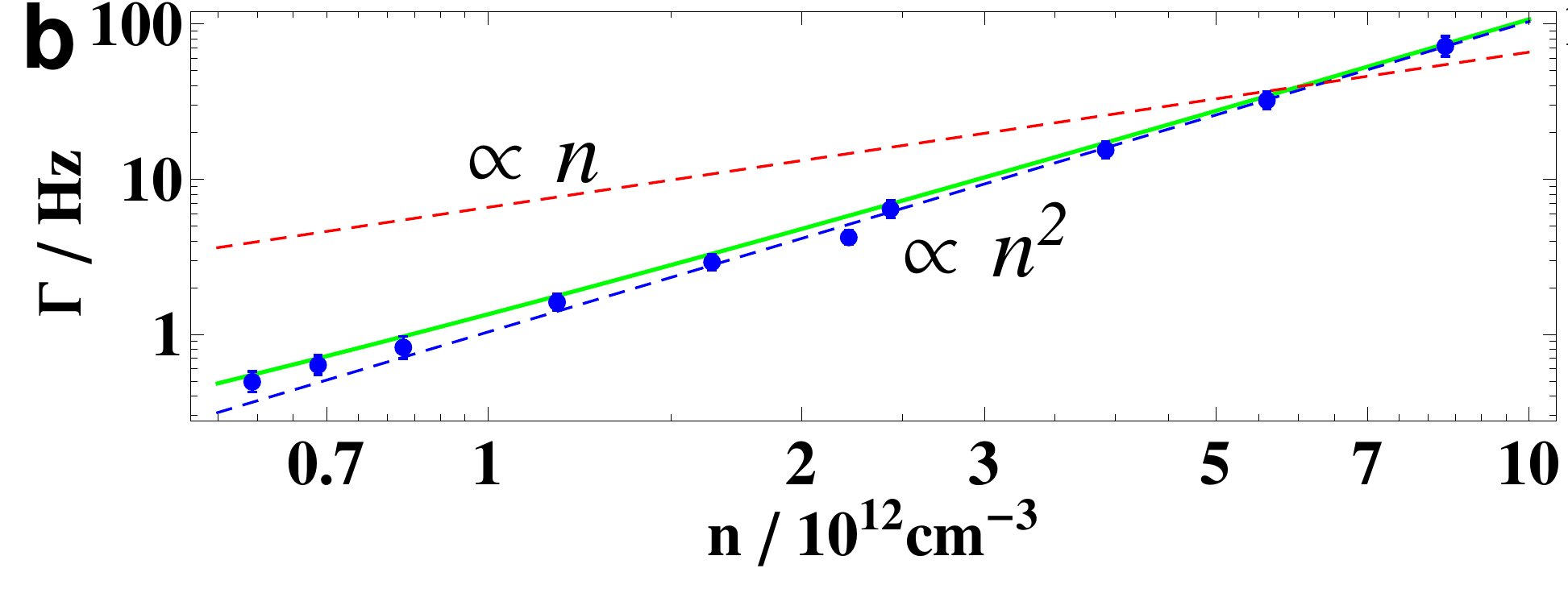}
\caption{(color online) \textbf{(a)} Logarithmic plot of $p$ as a function of the interaction time $t$ for 10 different initial atom peak densities. Each data point corresponds to an average of roughly 100 single ion experiments.
The dashed curves are simple exponential fits, while the solid curves originate from a simultaneous fit of Eq. \ref{eqn1} to the full data set  with two free parameters, the two-body rate coefficient $k_2$ and the three-body rate coefficient $k_3$ (see text for details).
\textbf{(b)} Double-logarithmic plot of the Ba$^+$ loss rates $\Gamma$ extracted from an exponential fit to each individual data set in (a) over the respective initial peak densities (filled circles).
A fit of the form $\Gamma=k_2 \cdot n + k_3 \cdot n^2$ to the loss rates yields a pure quadratic density dependence (blue dashed curve).
For comparison, this function was also plotted using $k_2$ and $k_3$ from (a) (green curve).
A pure linear dependence ($\Gamma \propto n$) does not describe the data (red dashed curve).
}\label{fig3}
\end{figure}

We are now ready to quantitatively analyze the reaction rate and to extract binary and ternary reaction rate constants. 
In order to obtain a high accuracy (and as a check for consistency) we perform \Ba lifetime measurements at 10 different initial peak densities (Fig. \ref{fig3}a).
Atom clouds with different densities are prepared by varying the trap frequencies and the atom number but keeping the atom temperature $T$ at a constant value of $T\approx 330 \:$nK.
This temperature  was chosen in order to be sufficiently above  the critical temperature $T_{\mathrm{c}}$ for Bose-Einstein condensation and to have negligible losses due to evaporative cooling.  
The trap depths are between  5 to 10 $\mu$K$\times k_{\mathrm{B}}$ resulting in trap frequencies of $(\omega_x , \omega_y , \omega_z)\approx 2\pi \times (16 \textrm{ to } 27;\:97 \textrm{ to } 151;\:107 \textrm{ to } 161)\:$Hz.
The mean ion kinetic energy is  4.0$\:(1.6)$ mK $\times k_{\mathrm{B}}$, as determined in \cite{Kruekow2016}.
The densities between $n\approx 6 \textrm{ to } 22 \times 10^{11}\:\textrm{cm}^{-3}$ are  prepared with $N\approx 40 \times 10^3$ atoms, while densities between $n\approx 24 \textrm{ to } 84 \times 10^{11}\:\textrm{cm}^{-3}$ are prepared with $N \approx 135 \times 10^3$ atoms.

In a first simple  analysis we do not include the density evolution and fit exponential decays (dashed lines) to each data set  in Fig. \ref{fig3}a.
The resulting loss rates $\Gamma$ are then plotted as a function of their respective initial atom densities $n(t = 0)$ in Fig. \ref{fig3}b.
By fitting $\Gamma=k_2 \cdot n + k_3 \cdot n^2$ (blue dashed line) we obtain a quasi-pure quadratic density dependence, where $k_3=1.03(2)(45)\times 10^{-24}\textrm{cm}^{6}\textrm{s}^{-1}$ and $k_2$ is consistent with zero.
For comparison,  if we try to describe the data only by two-body reactions,  $\Gamma  \propto  n $, no agreement is found (red dashed line).

Now, we perform a more rigorous analysis, where we account for the density decay during the interaction time, which can be as much as $20\:\%$ for  the experimental runs in Fig. \ref{fig3}.
This will enable us to also extract a reliable $k_2$ rate constant from the data.
With the previously described method  we determine  $n(t)$ for each  \Ba lifetime curve.
We then fit Eq. \ref{eqn1} to all of the 10 measured decays in Fig. \ref{fig3}a (solid lines) simultaneously, with only two free fit parameters, the binary and ternary rate coefficients $k_2$ and $k_3$, which amount to $k_2=3.1(6)(6)\times 10^{-13}\textrm{cm}^{3}\textrm{s}^{-1}$ and $k_3=1.04(4)(45)\times 10^{-24}\textrm{cm}^{6}\textrm{s}^{-1}$.
The first parenthesis denotes the $1\:\sigma$ statistical uncertainty of the fitted values.
The second one gives the $1\:\sigma$ systematic error due to the atom density uncertainty of $20\:\%$, which translates linearly into $k_2$ and quadratically to $k_3$.

Notably, both approaches yield the same $k_3$ within their uncertainties, but only the latter provides a non-zero $k_2$, which emphasizes the necessity to include the atom cloud decay.
We plot $\Gamma=k_2\cdot n + k_3 \cdot n^2$, using the extracted $k_2$ and $k_3$ coefficients in Fig. \ref{fig3}b (green curve).
Even at  low densities $n < 10^{12}$cm$^{-3}$ the green curve deviates only slightly from the pure three-body loss (blue dashed line), highlighting the small contribution of binary reactions to the total ion loss.

We now compare the obtained rate coefficients to the results of other groups in the field.
Our extracted two-body charge transfer rate coefficient $k_2$ for the \Ba + Rb system is compatible with a MOT measurement from \cite{Hall2013MP}
where an upper bound of $k_2 < 5 \times 10^{-13}\: \textrm{cm}^{3}\textrm{s}^{-1}$ is given for ground state charge-transfer.
An ab-initio calculation within \cite{Hall2013MP} predicts $k_2\approx 1 \times 10^{-14}\: \textrm{cm}^{3}\textrm{s}^{-1}$ which is a factor of 30 smaller compared to our present $k_2$.
A possible explanation for this large discrepancy is an additional two-body loss channel that might appear in our experiment.
 Indeed, calculated Ba$^+$-Rb molecular potential energy curves (see e.g. \cite{Hall2013MP}) indicate that the 1064$\:$nm dipole laser can near-resonantly photo-excite a colliding atom-ion pair to a repulsive molecular potential energy curve. 
For the potential curves that correlate with ionized Rb$^+$ and electronically excited neutral Ba, this process is experimentally indistinguishable from charge transfer.

We note that the three-body rate coefficient $k_3$, determined in this work for  \Ba + Rb + Rb is of similar magnitude as the one for  \Rbplus + Rb + Rb \cite{Haerter-2012} which is only by a factor of three smaller.
This can be understood as a consequence of the same long range atom-ion interaction potential of both systems, as it only depends on the polarizability of the Rb atom.
Indeed, a theoretical classical trajectory study predicted very similar three-body cross sections for \Ba and \Rbplus \cite{JPR-2015}.
Furthermore, since in cold reactive ternary collisions typically large, weakly-bound molecules should be formed \cite{Kruekow2016}, the short range details of the molecular interaction potential do not contribute.
This suggests a universal behavior of cold atom-atom-ion three-body recombination, leading to similar reaction rate coefficients for a variety of hybrid atom-ion systems.

In conclusion, we have studied reactive collisions of a cold, single \Ba ion in contact with an ultracold cloud of Rb atoms.
Mapping out the \Ba loss dependence on the Rb atom density enabled us to extract for the first time, both, the binary ($k_2$) and ternary ($k_3$) reaction rate coefficients at mK$\times k_{\mathrm{B}}$ ion energies.
The Ba$^+$-Rb-Rb three-body rate coefficient $k_3$ is comparatively large, about four orders of magnitude larger than the one for ultracold neutral Rb-Rb-Rb collisions  \cite{Esry1999}.
Moreover, it dominates  over the two-body loss down to comparatively low densities of $k_2/k_3\approx 3\times 10^{11} \textrm{cm}^{-3}$.
If working with degenerate quantum gases such as Bose-Einstein condensates with typical densities around $10^{14}\:$cm$^{-3}$, three-body recombination will occur on the sub-ms time scale, limiting the time for atom-ion experiments.
As shown in parallel work of ours \cite{Kruekow2016}, this time scale gets even shorter when lowering the collision energy  $E_{\mathrm{col}}$, as $k_3$ scales as $k_3 \propto E_{\mathrm{col}}^{-3/4}$.
In view of the number of proposed experiments where reactive collisions are unwanted, we expect a future demand for schemes to suppress three-body reactions besides the existing ones for two-body reactions \cite{Tomza-2015}.

We thank Olivier Dulieu, Jes\'{u}s P\'{e}rez-R\'{i}os and Chris H. Greene for fruitful discussions. This work was supported by the German Research Foundation DFG within the SFB/TRR21. A.K. acknowledges support from the Carl Zeiss Foundation.

\end{document}